\documentclass[useAMS,usenatbib]{mn2e}

\newcommand{\usco}{\mbox{U~Sco}}

\newcommand{\Msun}{\mbox{\,M$_\odot$}}

\newcommand{\vunit}{\mbox{\,km\,s$^{-1}$}}
\newcommand{\mic}{\mbox{$\,\mu$m}}
\newcommand{\pion}[2]{{#1}\,{\sc {#2}}}

\title[2010 outburst of U~Sco]{Near-infrared studies of the 2010 outburst of the
recurrent nova U~Scorpii}
\author[Banerjee et al.]{D. P. K. Banerjee$^1$\thanks{orion@prl.res.in},
R. K. Das$^1$, 
N. M. Ashok$^1$, 
M. T. Rushton$^2$, S. P. S. Eyres$^2$, \newauthor M. P. Maxwell$^2$, H. L.
Worters$^3$, A. Evans$^4$, B. E. Schaefer$^5$\\
$^1$Astronomy and Astrophysics Division, Physical Research Laboratory,
Navrangapura, Ahmedabad - 380009, Gujarat, India\\
$^2$Jeremiah Horrocks Insitute for Astrophysics and Supercomputing, University
of Central Lancashire, Preston, PR1 2HE, UK \\
$^3$South African Astronomical Observatory, PO Box 9, 7935 Observatory, South
Africa \\
$^4$Astrophysics Group, Keele University, Keele, Staffordshire, ST5 5BG, UK\\
$^5$Physics and Astronomy, Louisiana State University, Baton Rouge, LA 70803,
USA\\}

\usepackage{graphicx}
\begin{document}

\date{Accepted  Received }

\pagerange{\pageref{firstpage}--\pageref{lastpage}} \pubyear{2010}

\maketitle

\label{firstpage}

\begin{abstract}
We present near-IR observations of the 2010 outburst of \usco. $J\!H\!K$
photometry is presented on ten consecutive days starting from 0.59 days after
outburst. Such photometry can gainfully be integrated into a larger database of
other multi-wavelength data which aim to comprehensively study
the evolution of \usco. Early near-IR spectra, starting from 0.56 days after
outburst, are presented and their general characteristics discussed. 
Early in the eruption, we see very broad wings in several spectral lines, with
tails extending up to $\sim10\,000$\vunit\ along 
the line of sight; it is unexpected to have a nova with ejection
velocities equal to those usually thought to be exclusive to supernovae.
From recombination analysis, we estimate
an upper limit of $10^{-4.64^{+0.92}_{-0.74}}\Msun$ for the ejected mass.
\end{abstract}


\section{Introduction}
The well known recurrent nova (RN) U Scorpii has undergone at least
six previous outbursts, in 1863, 1906, 1936, 1979, 1987 and 1999 and
search of archival data resulted in the detection of three additional
outbursts, in  1917, 1945 and 1969 \citep{b28}. Its latest outburst,
on 2010 January 28.4385 UT, was discovered by B. H. Harris and S. Dvorak
\citep{b29b}.

The latest outburst was predicted to occur around year $2009.3\pm1$
\citep{b27}, based on the average brightness and time between eruptions.
The binary components of \usco\ consist of a massive white
dwarf (WD) and a low mass companion in a 1.2305631 day period eclipsing
system \citep{b26}. Though the outburst of RNe and classical novae
share a common origin in a thermonuclear runaway on a WD surface that has
accreted matter from a companion star, an important distinguishing
feature in RNe is the smaller amount of accreted mass required and
consequently the shorter intervening period to trigger the outburst. Thus
RNe are well suited to provide observational inputs and constraints to
nova trigger theories \citep{b27}. RNe are also of particular interest as
they are possibly progenitors of Type Ia supernovae \citep{b33, b15}.

The last three outbursts of \usco\ were well studied, especially in the optical
\citep{b7, b35, b32, b2, b19, b16}. The only major IR study of \usco\
was during the 1999 outburst by \cite{b13}, who obtained spectra between 2.34
and 27.28 days after outburst. An X-ray study of the super-soft phase was also
made by \cite{b17} for the 1999 eruption. These studies aimed at determining
important physical parameters like the mass of the ejecta, spectral type of the
secondary, and estimating the He abundance among other parameters. The present
outburst was widely anticipated and a major world-wide multi-wavelength campaign
was planned well in advance. As a result extensive data have been
collected and preliminary results have been reported in the UV and X-rays from
SWIFT \citep{b23, b29, b30, b31} and CHANDRA observations \citep{b22}, in the
optical \citep{b1} and the infrared \citep[IR;][]{b4, b12}. In this paper we
present near-IR spectroscopic and photometric data during the early decline
phase.
The spectra taken 0.59 days after outburst are the earliest to be recorded for
this object in the near-IR.

\section{Observations}
\subsection{Mt. Abu}
Near-IR observations were carried out in the $J\!H\!K$ bands at the Mt.
Abu 1.2m telescope in the early declining phase of the outburst. 
The comparison star for photometry was SAO\,159825 ($J=8.26, H =7.92,
K=7.88$). Spectra in the wavelength range 1.09--2.2\mic\ were obtained from
day 0.56 to day 4.55 at a resolution of $\sim1000$ using a Near-Infrared
Imager/Spectrometer with a $256\times256$ HgCdTe NICMOS3 array. Spectral
calibration was done using OH sky lines and telluric features that register with
the stellar spectra. $\omega^1$~Sco (B1V, $T_{\rm eff} = 25400$~K) was chosen as
the standard star and observed at similar airmass to \usco\ to ensure the
ratioing process removes telluric lines. Subsequent reduction of the spectra and
processing of the  photometric data follow a standard procedure that is
described for e.g. in \cite{b21}. All data reduction was done using IRAF tasks.

\vspace{-3mm}

\subsection{ESO}
IR spectroscopy was obtained at the 3.6m New Technology Telescope (NTT), using
the SOFI IR spectrograph and imaging camera \citep{sofi}. Data were obtained
on days 5.41 and 9.43 using the blue and red low resolution grisms, giving
a wavelength coverage of 1--2.5\mic\ at $R\sim1000$. Flux calibration and the
removal of atmospheric features was achieved by the dividing the target spectra
by the spectra of the standard star HIP 45652 (B9V) on 2010 February 3
(MJD~5230.6).
The data were wavelength calibrated using a Xenon lamp. A log of
photometric and spectroscopic observations is given in Tables~\ref{phot} and
\ref{spec} respectively.

\begin{table}
\caption{
Photometry of U Sco from Mt. Abu. The mid-time of observations are given in
MJD; $\Delta{t}$ is time since outburst, taken to be MJD\,5224.9385 
\citep[2010 January 28.4385 UT;][]{b29b}.\label{phot}}
\begin{tabular}{cccccc}
\hline \\
MJD & $\Delta{t}$ (d) & $J$ & $H$ & $K$ \\\hline  \hline \\
5225.5286  &0.59&7.00$\pm$.01 & 6.72$\pm$.02 &6.32$\pm$.01 \\
5226.5189  &1.58&8.05$\pm$.02 & 7.88$\pm$.05 &7.33$\pm$.06 \\
5227.5402  &2.60&8.72$\pm$.06 & 8.67$\pm$.06 &8.09$\pm$.04 \\
5228.5349  &3.60&9.13$\pm$.12 & 9.38$\pm$.11 &  8.60$\pm$.10 \\
5229.5360  &4.60&9.79$\pm$.09 & 9.82$\pm$.10 &  9.27$\pm$.15 \\
5230.4666  &5.53&10.24$\pm$.11 & 10.25$\pm$.12& 9.65$\pm$.08 \\
5231.4829  &6.54&11.06$\pm$.20 & 10.61$\pm$.17 &9.84$\pm$.38 \\
5232.4563  &7.52&11.38$\pm$.14 & 11.23$\pm$.24 &10.23$\pm$.59\\
5233.4805  &8.54&12.11$\pm$.15 & 12.06$\pm$.29 &11.07$\pm$.33 \\
5234.4823  &9.54&12.53$\pm$.21 & 11.97$\pm$.19 &11.82$\pm$.47 \\\hline
\end{tabular}
\end{table}

\begin{table}
\caption{Spectroscopy of U Sco. Integration time in s; for
blue and red grisms for SOFI.\label{spec}}
\begin{tabular}{ccccccc}
\hline \\
MJD & $\Delta{t}$ (d) & Site & $J$ & $H$ & $K$ \\\hline  \hline \\
5225.4959 &0.56 &MtA & 120 & 120 & 150 \\
5226.4820 &1.54 &MtA & 200 & 180 & 180 \\
5227.4886 &2.55 &MtA& -- &  -- & 250 \\
5228.4850 &3.54 &MtA& 500 & -- & 500 \\
5229.4845 &  4.55 & MtA &600 & 500 & 500 \\
5230.3476 &  5.41 & ESO &360 &  & 480 \\
5234.3670 &  9.43 & ESO& 720 &  & 960 \\ \hline
\end{tabular}
\end{table}

\begin{figure}
\centering
\includegraphics[bb= 84 316 345 576,
width=3.0in,height=3.0in,clip]{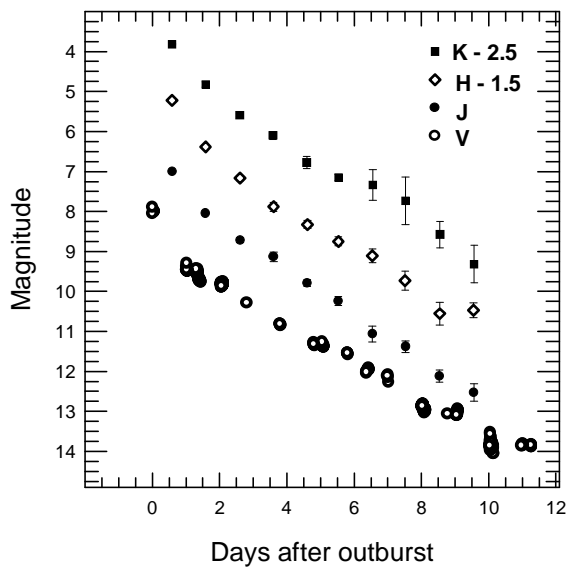}
\caption[]{ The near-IR and visual light curve of U Sco during the early decline
phase. The $V$ band data are from AAVSO. The $J, H, K$ magnitudes have been
offset by requisite amounts (as indicated in the figure) for clarity.}
\label{fig1}
\end{figure}


\begin{figure*}
\centering
\includegraphics[bb=61 376 554 691,width=7.0in,height=4.3in,clip]{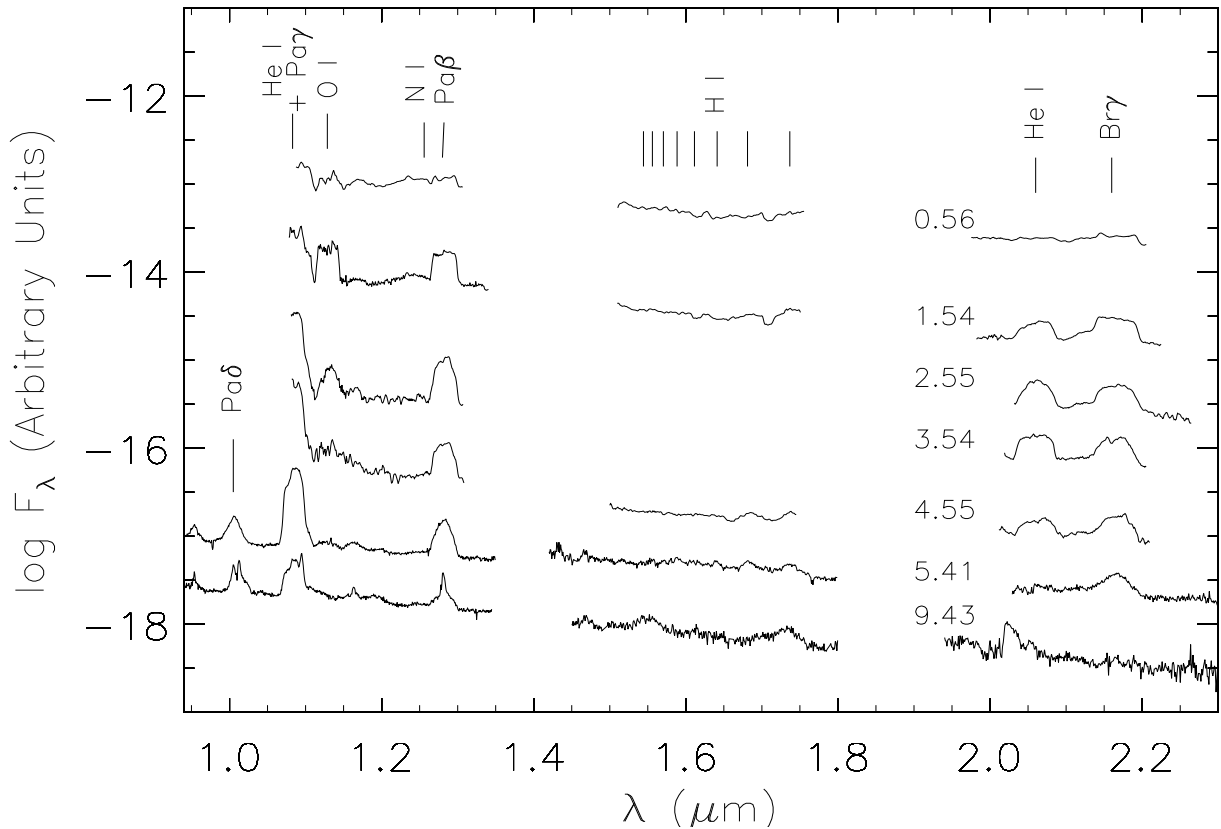}
\caption[]{\label{fig2} The $J\!H\!K$ spectra; the time (in days) from outburst
is given next to the $K$-band data.}
\end{figure*}


\section{Results}
\subsection{IR light curves and spectra}
The $J\!H\!K$ lighcurves are shown in Fig.~\ref{fig1}, along with a matching
portion of the optical lightcurve for comparison. The evolution of the near-IR
lightcurve is rather similar to the optical. Since sampling is only on a daily
basis, we have interpolated between points to obtain mean ($J,H,K$ averaged)
$t_{2}$ and $t_{3}$ times of 2.4 and 4.0~days respectively for the near-IR.
However these values are likely to be on the high side as we missed the fast
declining stage between time zero to 0.59 days (when our first data point was
recorded). For the present outburst, \cite{b20} estimate $t_{2} = 1.8$ and
$t_{3} = 4.1$~days in the $V$ band.
The distance to \usco\ as based on $t_2$ and $t_3$ has been presented in a
unified manner by \cite{b28} to be 37.7~kpc, but this is superceded by the
blackbody distance of the companion star during the total eclipse, which is
$12\pm2$~kpc \citep{b28}.

The spectra are presented in Fig.~\ref{fig2}.
The spectra are similar to those seen in novae outbursts occuring on
massive white dwarfs, examples of which are V597~Pup, V2491~Cyg and RS~Oph
\citep{b21, b6}. The prominent features detected in \usco\ are the
\pion{He}{i} 1.0830\mic\ and Pa$\gamma$ 1.0938\mic\ lines (blended),
\pion{O}{i} 1.1287\mic, Pa$\beta$ 1.2818\mic, \pion{He}{i} 2.0581\mic\ and the
\pion{H}{i} Brackett series lines in the $H$ band. The Brackett series lines in 
the $H$ band are severely blended due to the large line widths.

A feature at $\sim1.163$\mic\ is likely \pion{N}{i} 1.1625, 1.1651\mic. While 
this feature also coincides with \pion{He}{ii} 1.16296, 1.1676\mic, we
believe that \pion{N}{i} is the more the likely: if it is
\pion{He}{ii}, its strength is expected to increase with time, as the level of
ionization and excitation increases. While such behaviour is seen for the
\pion{He}{i} 1.083 and 2.0581\mic\ lines (Fig.~\ref{fig2}), it is not seen for
the 1.163\mic\ feature.

U~Sco does not seem to show the presence of prominent carbon lines, and in
general C lines are weak in optical spectra during outburst \citep{b7,RI}.
Carbon emission is a defining IR signature of novae occurring on CO white dwarfs,
with mass $\la1.2$\Msun\ (the so-called ``CO novae''). Typical spectra of CO
novae, and their differences from the present spectra, can be seen in the cases of
V2274~Cyg \citep{b25} and V1280~Sco and V2615~Oph \citep{b10,b11}. For example,
in V2274~Cyg the \pion{C}{i} line at 1.44\mic\ ($^3$P$-^3$D) was comparable in
strength with Br$\gamma$ but there is no evidence for it in U~Sco.

The emission lines are remarkably broad and the \pion{H}{i} lines (Pa$\beta$ and
Br$\gamma$) consist of a core component and possible broad wings, a detailed
discussion of which is given in Section~\ref{fast}. The core component has a
FWZI in the range of 9\,000--10\,000\vunit\ for all the prominent lines. A
triple-peaked profile is seen in the Pa$\beta$, Br$\gamma$ and
\pion{He}{i}2.058\mic\ lines in the earliest spectra, 0.56 days after
outburst which, however, disappears by the next day (Figs~\ref{fig2},
\ref{fig3}). Similar triple-peaked (``Batmanesque'') structure was seen in
H$\alpha$ in early optical spectra \citep{b3}. A triple-peaked profile is also
seen in the \pion{O}{i} 1.1287\mic\ line on 2010 January 28.996 UT, but we
caution that the region around this line has low atmospheric transmission and
artificial structures can be generated in the profile during spectrum
extraction.

The line fluxes for day 0.56 are given in Table~\ref{fluxes}.

\begin{table*}
\caption{Line fluxes for day~0.56, dereddened for $E(B-V)=0.2$. A full version of
this table is avalable in the on-line version. \label{fluxes}}
\begin{tabular}{lccccccc}
 Line &  OI 1.1287 & NI 1.1625,1.1651 & NI 1.2461,1.2470 & P$\beta$ 1.2818 & Br11
 1.6806 & HeI 2.0581 & Br$\gamma$ 2.1655 \\ \hline
 Flux ($10^{-18}$~W~cm$^{-2}$)  & 5.80 & 3.43 & 4.50 & 4.10 &  1.50 & 0.59 & 2.25\\
 Uncertainty ($10^{-18}$~W~cm$^{-2}$)  & $\pm0.17$ & $\pm0.10$ & $\pm0.8$ &
 $\pm0.18$ &  $\pm0.03$ & $\pm0.04$ & $\pm0.08$ \\
\hline \\
\end{tabular}
\end{table*}

\vspace{-3mm}

\subsection{Evidence for high-velocity ejecta\label{fast}}
\begin{figure}
\centering
\includegraphics[bb=104 392 379 666,width=3.0in,height=3.0in,clip]{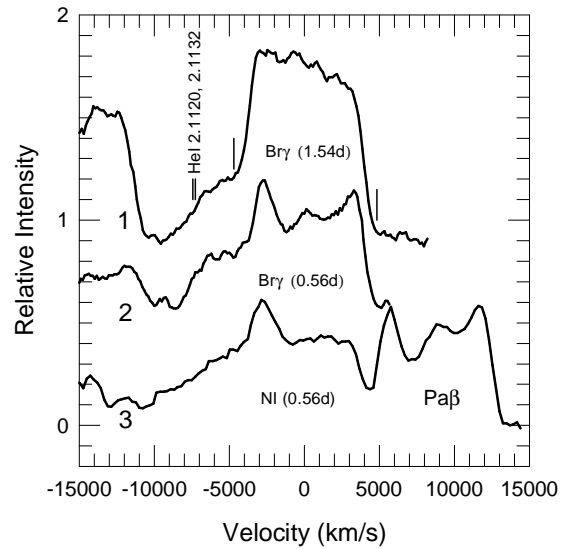}
\caption[]{Velocity profiles 1 and 2 are for Br$\gamma$ on days 1.54 and 0.56
respectively; they have a core component between $-4700$ and
$+4850$\vunit\ (marked by lines), and an extended blue wing (see
Section~\ref{fast}). The expected positions of \pion{He}{i} 2.1120, 2.1132\mic\
lines are shown. Profile 3 is for the \pion{N}{i} 1.2461, 1.2470\mic\ line on
day 0.56, which also shows an extended  blue wing. The ordinate is
in arbitrary units with the profiles offset for clarity.}
\label{fig3}
\end{figure}

The {\em FWHM} of the IR emission lines in the 1999 eruption indicated
velocities of $\sim2\,500$\vunit\ \citep{b13} and {\em FWZI} of
$\sim9\,500$\vunit\ over the first 5~days; likewise the {\em FWZI} of the Balmer
lines were $\sim10\,000$\vunit\ in the 1979 \citep{b7} and 1999 \citep[][data
obtained on day~0.65, close to our first spectrum]{b16}
eruptions. Examination of the Br$\gamma$
profile in the 2010 eruption (Fig.~\ref{fig3}) shows a core component with a
{\em FWZI} of $\sim9\,500\vunit$. More interesting however is the very extended
blueward wing in the profiles for the first 2 days, which extend to about
10\,000\vunit\ from the line center. It is not clear whether an equivalent red
wing exists for Br$\gamma$, as our spectra do not extend that far redward. For
the present we concern ourselves with the 10\,000\vunit\ blueward wing.

We first establish that the extended blue wing of Br$\gamma$ is genuine and
intrinsic to the line by noting that the \pion{N}{i}~1.2461, 1.2470\mic\
lines, also plotted in Fig.~\ref{fig3}, exhibit a similar wing. It is
difficult to conclude whether other lines have similar wings. Pa$\beta$ and
\pion{O}{i}~1.1287\mic\ are closely flanked on the blue side by other lines.
There may be a wing on \pion{He}{i} 2.0581\mic\ as there is a small undulation
at $\sim2.006$\mic, $\sim-7500$\vunit, for first 2 days (see
Fig.~\ref{fig2}); a similar ``bump'' is seen in the 1999 spectrum of
\cite{b13}. But we again caution that the 
position of this undulation is in a region of poor atmospheric transmission.
Could the wings in Fig.~\ref{fig3} be caused by additional spectral lines? No
such line is expected in the case of the \pion{N}{i}, but the
Br$\gamma$ wing is the site of the \pion{He}{i}~2.1120, 2.1132\mic\ lines.
However the expected positions of these \pion{He}{i} lines, marked in
Fig.~\ref{fig3}, are not too well centered with the extended wing. Thus it is
unlikely that they contaminate the wing. Additionally the \pion{He}{i}
2.1120\mic\ line may be expected to be weaker by a factor of at least 20
compared to \pion{He}{i}~2.0581\mic\ \citep{b8}. Observations of other novae
\citep[e.g. RS Oph;][]{b6} strongly support this. However, Fig.~\ref{fig3}
shows that the strength of the Br$\gamma$ blue wing, especially in the spectrum
0.56 days after outburst, is too strong compared to \pion{He}{i} 2.0581\mic,
for it to be caused by the \pion{He}{i} 2.1120\mic\ line. We thus consider it
unlikely that there is any significant presence of \pion{He}{i} 2.1120\mic\ and
conclude that the broad wing is intrinsic to Br$\gamma$ and therefore
suggestive of material moving at $\sim10\,000$\vunit; even if this represents
the line-of-sight velocity of ejected material, it is well in excess of
expected ejecta speeds, even for RNe.

It is possible that this material arises in a bipolar flow. Such bipolar flows
have been observed in novae \citep[e.g. RS~Oph, V445~Pup;][]{b5,b36}, and
are explained on the basis of ejected material encountering density
enhancements in the equatorial (orbital) plane compared to the polar direction;
the outflowing material thus expands more freely in the polar direction,
leading to high velocity polar flows. Such an outflow would be expected to be
perpendicular to the orbital plane and, in the case of \usco\ \citep[as it is
an eclipsing binary with inclination angle $\sim80^\circ$;][]{b14}, close to
the  plane of the sky. If we assume inclination $\sim80^\circ$ and opening
angle $\sim15^\circ$ for the jet, the 10\,000\vunit\ line-of-sight velocity
translates to a space velocity of $\sim23\,000$\vunit. Either way a very fast
flow is being witnessed, possibly the fastest seen in any nova eruption.

We note that most of the spectral studies of the 1999 outburst showed the
H$\alpha$ profile to have similar broad wings. The doubt again arises whether
these are intrinsic to the H$\alpha$ line or caused by additional lines. For
example \cite{b16}, in a spectrum taken 16 hrs after maximum,
assigns the \pion{N}{ii} line at 6482\AA\ as a possible
cause for the extended blue wing of the H$\alpha$ profile. However, if it is an
intrinsic structure and not really \pion{N}{ii} 6482\AA\ which is contributing,
then it is seen that the wing extends to about 10\,900\vunit\ from the line
center \citep[Figure 2 of][]{b16}. This is in good agreement with the above
discussion, and is supporting evidence for a very fast flow.

\subsection{Mass estimate using recombination analysis\label{mass}}
In the analysis below we are able to set a useful upper limit on the ejecta
mass; an exact determination is difficult due to uncertainty in the conditions
in the ejecta, and the range of velocities present.
For the recombination analysis we use the data for 2010 February 2.8
(day~5.41); we consider this to be the most favourable epoch, because the
lines are more likely to be optically thick at earlier times, while the hydrogen
lines for day~9.43 are likely contaminated by \pion{He}{ii} Pickering lines,
which occur at the same wavelengths as the \pion{H}{i} lines. We
use the best-defined lines from our data for this epoch, namely Br$\gamma$,  
Pa$\beta$ and Br11 (1.6806\mic).
The strengths of the other Brackett series lines in the $H$ band are difficult
to assess because of blending. We assume there is no significant contribution to
the strength of these \pion{H}{i} lines from the \pion{He}{ii} Pickering lines
on day~5.41.

We proceed on the assumption that the ejecta are optically thin in the
Pa$\beta$, Br$\gamma$ and Br11 lines. It is expected that nova ejecta have
density in the range $n_{\rm e} = 10^{10}$ to $10^{12}$~cm$^{-3}$ in the
early stages. Thus, for example, for $n_{\rm e} =
10^{11}$~cm$^{-3}$ and $T=10^{4}$~K, Case B predicts 
ratios of 4.76 and 3.8 for Pa$\beta$/Br$\gamma$ and Br$\gamma$/Br~11
respectively. The observed ratios in \usco\ for Pa$\beta$/Br$\gamma$
($4.29\pm0.15$) and Br$\gamma$/Br~11 ($2.69\pm0.80$) are reasonably in agreement
with these predictions, although there is no plausible combination of $n_{\rm
e}$ and $T_{\rm e}$ that has the Pa$\beta$/Br$\gamma$ ratio as low as
$\sim4.3$ \citep[the 
Case B emissivities at $n_{\rm e} = 10^{11}$~cm$^{-3}$, $T=10^{\rm 4}$~K are
$3.1\times10^{-26}$, $6.5\times10^{-27}$ and $1.7\times10^{-27}$~erg~cm$^{3}$
s$^{-1}$ for the Pa $\beta$, Br $\gamma$ and Br~11 lines respectively;][]{b34}. 
If the lines under consideration are largely optically thick, considerable
changes are seen in the ratios \citep[especially in the the Br$\gamma$/Br 11
ratio which can even drop below unity; see][]{b6}.

The mass of the emitting gas is given by
\[ M = \sqrt{4\pi D^2 \,\, m_{\rm H}^2 \,\, (fV/\epsilon)} \]
where $D$ is the distance, $m_{\rm H}$ the proton mass, $f$ the observed
flux in a particular line, $\epsilon$ the corresponding case B
emissivity; $V$ is the volume of the emitting gas, which is
$(\frac{4}{3} {\pi}[v\,t]^3 \, \phi)$ where $\phi$, $v$ and $t$ are the
filling factor, velocity and time after outburst respectively. We use $D =
12\pm2$~kpc \citep{b28}, $v = 5\,000$\vunit\ (although a range of a factor
2 either side of this value is implied by our data) and values of $f = (7.31 \pm
0.01)\times10^{-12}$, $(1.70\pm 0.06)\times10^{-12}$ and $(6.3\pm
0.2)\times10^{-13}$~ergs~s$^{-1}$~cm${^{-2}}$ measured for the core components
of the Pa $\beta$, Br$\gamma$ and Br11 lines respectively.

The greatest uncertainty in our analysis arises from our ignorance of the
electron density and temperature in the ejecta, and from our assumption
that Case~B applies. If we suppose that
$10^8\le{n_e}\mbox{(cm$^{-3}$)}\le10^{12}$ and
$10^4\le{T_e}\mbox{(K)}\le3\times10^4$, then the mean $\log\epsilon$ (in
erg~s$^{-1}$~cm$^{3}$) is $-25.90^{+0.59}_{-0.32}$, $-26.68^{+0.55}_{-0.35}$ and
$-27.23^{+0.46}_{-0.37}$ \citep{b34} for Pa$\beta$, Br$\gamma$ and Br11
respectively, where the ``error bars'' represent the range of values.
Over this range of $n_{\rm e}$ and $T_{\rm e}$ the Case~B ratios for
P$\beta$/Br$\gamma$ and Br11/Br$\gamma$ range from 4.8 to 8.4, and 2.7 to 4.5
respectively. As already noted, the observed P$\beta$/Br$\gamma$ ratio is less
than the lowest value expected for Case~B, possibly indicating that
Pa$\beta$ may not be optically thin; the mass derived from Pa$\beta$ may
therefore be an underestimate.
We find $\log{M}=10^{-4.71^{+0.55}_{-0.49}}\,\phi$\Msun, 
$\log{M}=10^{-4.64^{+0.53}_{-0.24}}\,\phi$\Msun\ and 
$\log{M}=10^{-4.58^{+0.51}_{-0.50}}\,\phi$\Msun, 
from Pa$\beta$, Br$\gamma$ and Br11 respectively. 
The errors in $f$, $D$, $v$ and the uncertainties in $\epsilon$ have been added
in quadrature, although we recognise that this is not necessarily robust (e.g.
the errors are asymmetric \cite[see][for a discussion of this point]{barlow},
while the uncertainties in $\epsilon$ are not in any sense ``errors'' and are
not distributed normally). Our best estimate of the ejecta mass is therefore
$10^{-4.64^{+0.92}_{-0.74}}\,\phi$\Msun\ ($\sim2.2\times10^{-5}\,\phi$\Msun);
since we must have that $\phi<1$, $M<10^{-4.64^{+0.92}_{-0.74}}$\Msun.

An alternative estimate of the ejected mass can be provided by free-free
(f--f) emission \citep[cf.][who found
$M\sim\mbox{a~few}\times10^{-7}$\Msun]{b13}. We constructed the SED using $V$
magnitudes from AAVSO, our near-IR magnitudes, and reddening $E(B-V)$ in the
range 0.2 to 0.56 \citep{b7, b14}. We have searched for a f--f excess 
on the first six days, where the errors on the $J\!H\!K$ magnitudes are small.
Depending on the reddening, blackbody fits with effective temperatures in
the range 6\,000 -- 8\,000~K reasonably fit the data. However, while our
analysis shows that a f--f excess might be present in the data, especially
in the $K$ band, it is marginal and difficult to quantify. In view of
this we do not use f--f to estimate a mass. Observations at mid-IR
wavelengths should reveal any f--f emission as the emissivity is proportional to
$\lambda^{2}$.

\section*{Acknowledgements} The research at PRL is funded by the Dept. of
Space, Government of India. We acknowledge with thanks the use of AAVSO data.

\label{lastpage}

\end{document}